# Tunneling conductivity in anisotropic nanofibre composites: a percolation-based model

Avik P. Chatterjee[*,a]

and

Claudio Grimaldi[b]

[a]Department of Chemistry, SUNY College of Environmental Science and Forestry, One Forestry Drive, Syracuse, New York, U.S.A. 13210

[b]Laboratory of Physics of Complex Matter, Ecole Polytechnique Fédérale de Lausanne, Station 3, CP-1015 Lausanne, Switzerland

[*] Corresponding author: E-mail: apchatte@esf.edu ; Fax: 315-470-6856

**Abstract**

The Critical Path Approximation ("CPA") is integrated with a lattice-based approach to percolation to provide a model for conductivity in nanofibre-based composites. Our treatment incorporates a recent estimate for the anisotropy in tunneling-based conductance as a function of the relative angle between the axes of elongated nanoparticles. The conductivity is examined as a function of the volume fraction, degree of clustering, and of the mean value and standard deviation of the orientational order parameter. Results from our calculations suggest that the conductivity can depend strongly upon the standard deviation in the orientational order parameter even when all the other variables (including the mean value of the order parameter $\langle S \rangle$) are held invariant.

## 1. Introduction:

The formation of an infinite network of connected (in the sense of spatial proximity) particles under appropriate conditions of composition and state of dispersion is referred to as percolation, and is frequently correlated with large variations in properties (such as the conductivity, denoted $\sigma$) of composite materials [1,2]. Analytical theory based upon both lattice [3] and continuum representations [4-8] as well as computer simulation studies [9-12] reveal that the volume fraction occupied by the particles at the percolation threshold, denoted $\phi_c$, is strongly dependent upon the particle shape. For the case of elongated, rod-like, electrically conductive nanoparticles that can be approximately modeled as cylinders, theory and experiment suggest that the conductivity of a composite based upon such filler species also depends upon the degree of orientational alignment of the particles [13-16]. The extent of uniaxial orientational ordering is customarily quantified through the averaged value of the Legendre polynomial of second order, denoted $\langle S \rangle = \langle P_2(\cos(\theta)) \rangle$, where $\theta$ represents the angle between the longitudinal axis of a selected particle and the direction of alignment. Computer simulations hint that the typical result to be expected is for the conductivity to decrease with increasing values of $\langle S \rangle$, which is partially confirmed by experiments even though non-monotonic variations in $\sigma$ as a function of the orientational order parameter have also been reported [13]. A recent analysis [17] that employed the Effective Medium Approximation ("EMA") provided an estimate for the anisotropy in the effective tunneling conductance between a pair of rod-like particles, that is, for the dependence of the conductance upon the relative angle between the longitudinal axes of the particle pair.

This anisotropy (the mutual conductance was shown to be larger for parallel as opposed to perpendicular orientations) was explicitly accounted for in the EMA-based investigation.

Recently, a heuristic approach to describing percolation by elongated objects has been developed [3,18] that exploits an analogy between continuum and lattice representations, which are linked by equating the average number of contacts per particle (in the continuum problem) to the mean number of nearest-neighbor occupied sites (in the lattice problem). This formalism has been applied to examine how $\phi_c$ depends upon polydispersity in the aspect ratios of the particles for isotropic systems [3,18], and also upon the order parameter $\langle S \rangle$ in polydisperse, partially aligned systems [19]. The impact of deviations from a perfectly random particle arrangement, such as effects due to clustering, can be addressed through an extension of this approach [18] (albeit one that should be viewed as being strictly operational). A pathway has been explored recently [20] that combines this model with the Critical Path Approximation [21] ("CPA") to describe the variation in conductivity with particle volume fraction for an isotropic orientational distribution.

The present work reports upon a generalization that includes the effects of orientational ordering, as well as anisotropic tunneling conductance between particle pairs, upon the conductivity within a CPA and percolation-based framework. A (perhaps somewhat unorthodox) choice for the orientational distribution function ("ODF") is adopted that permits varying the mean value and standard deviation in the order parameter independently, alongside parameters that operationally characterize non-randomness in the distribution of the particle centers. Our theory, which includes approximations that are introduced in the interests of constructing a model that is nearly entirely analytical in nature, is developed in Section 2. Results from calculations performed using this model are discussed, and concluding remarks are presented, in Sections 3 and 4, respectively.

**2. Model and Theory:**

**2.1. Anisotropic tunneling conductance and the critical path approximation for partially aligned rod systems:**

We consider a dispersion of monodisperse, rigid, cylindrical, rod-like conducting particles within an insulating matrix. The particle cores are assumed to have uniform radii and lengths, denoted $R$ and $L$, respectively, and it is assumed that $L$ is far greater than $R$ $(L/R \gg 1)$. The particle volume fraction (denoted $\phi$) equals: $\phi = \pi \rho R^2 L$, where $\rho$ represents the number of rods per unit volume. An expression for the effective tunneling conductance between a pair of rods in such a system has been derived in Ref. [17], and forms the starting point for the present work. For a configuration such that the centers of each member of the pair of rods are both located on the line of shortest distance between the rod axes (figure 1 of Ref. [17]), equations (22) and (27) of Ref. [17] yield the following result for the rod-to-rod tunneling conductance:

$$\frac{\sigma}{\sigma_0} = \frac{e^{-4\lambda/\xi}}{\left[\dfrac{4\pi R \xi}{L^2} + \sin^2(\gamma)\right]}, \qquad (1)$$

where $\gamma$ denotes the angle between the longitudinal axes of the rods, $2\lambda$ is the distance of closest approach between the surfaces of the pair of particles, and $\sigma_0$ represents a prefactor for the conductance. In equation (1), $\xi$ is the length scale associated with the decay of electron tunneling between the particle surfaces. The value of $\xi$ depends upon the specific system under consideration, but has been estimated to typically be within the range [22]: 0.1 nm < $\xi$ < 10 nm. (In the interests of completeness, it should be noted that the result in equation (27) of Ref. [17] also includes dependence upon the lateral separation between the rod centers, denoted by the variable $|z_{ij}|$ in that work. Figure 2 of Ref. [17] reveals, however, that the dependence upon $|z_{ij}|$ is quite weak for most of the range of $\gamma$. We adopt the simplifying assumption of setting: $|z_{ij}| = 0$, which leads to equation (1) of the present report). For subsequent use in conjunction with the Critical Path Approximation [21] ("CPA"), equation (1) can be rewritten to express the separation between particle surfaces in terms of the tunneling conductance and the angle between the rod axes:

$$\frac{\lambda(\gamma)}{R} = -\frac{1}{4}\left(\frac{\xi}{R}\right)\left[\ln\left(\frac{\sigma}{\sigma_0}\right) + \ln\left(\frac{4\pi R\xi}{L^2} + \sin^2(\gamma)\right)\right]. \tag{2}$$

Equation (2) shows that for a fixed relative angle $\gamma$, $\lambda/R$ is a monotonically decreasing function of $\sigma/\sigma_0$, which is physically reasonable.

We next adopt the CPA, along with a lattice-based approach to percolation, to estimate the conductivity for a composite characterized by distributions over the relative angles and distances of separation between the rods as well as by varying degrees of (uniaxial) orientational order. For a given arrangement of particles, calculation of the conductivity by way of the CPA proceeds through the following operational steps [21]: (i) a conductance is assigned to each pair of rods; (ii) each of these pairwise conductances is next set to zero/replaced by an infinite resistance; (iii) non-zero conductances are then restored between pairs of rods, from zero upwards through progressively increasing values of $\sigma/\sigma_0$, until eventually: (iv) a percolating network is formed among those rod pairs that are connected through non-zero conductances at a threshold value for $\sigma/\sigma_0$. The critical threshold value of $\sigma/\sigma_0$ at which such a percolating network of connections first arises provides the desired estimate for the conductivity of the composite. This methodology has been applied in recent investigations into the conductivity of composites [20,23], and enables integrating ideas from percolation theory with estimates (such as that in equation (2)) for how the conductance between a pair of particles varies with their separation and relative angular displacement.

In this work, equation (2) will be used as a criterion for pairwise connectedness between particles with a surface separation that does not exceed $2\lambda(\gamma)$, where this criterion maps into an equivalent effective conductivity. The value of $\sigma/\sigma_0$ that implies a percolation threshold in the connectedness sense when equation (2) is used to assign the range over which pairs of particles are defined as being "connected" yields our estimate for the conductivity of the composite. Calculation of the percolation threshold for partially aligned systems of rods in the context of the $\gamma$-dependent connectedness criterion that is implied by equation (2) is discussed in Section 2.2.

**2.2. Lattice-based estimation for the percolation threshold:**

We next modify a heuristic mapping between lattice and continuum percolation problems that has been applied to investigate systems of rod-like particles to analyze the situation of present interest. The mapping in question is described in Refs. [3] and [18], and has previously been extended to treat polydisperse systems of partially oriented rods [19]. This method posits an analogy between rods of varying aspect ratios and occupied sites (or vertices) of varying degree on a tree-like lattice; vacant sites on the lattice represent locations to which particles are added as the volume fraction is increased. Contacts between particles are represented by links (or edges) joining adjacent occupied sites in the lattice model. The site occupation probability is equated to the volume fraction occupied by the particles in the continuum problem, and the degrees (or co-ordination numbers) for the vertices are assigned by enforcing equality of the number of pairwise, nearest-neighbor contacts for a given criterion for connectedness.

We begin by considering pairs of rods each of length $L$ and radius $R$, where $\gamma$ represents the angle between the longitudinal axes and the particles are defined as being "connected" if the smallest distance between their surfaces is less than $2\lambda(\gamma)$. For a random spatial distribution of the particle centers, the leading-order contribution to the average number of contacts (denoted $n_c$) that an individual rod experiences with all of the other particles is:

$$n_c \cong \left(\frac{4\phi}{\pi}\right)\left(\frac{L}{R}\right)\left\langle\left(\frac{\lambda(\gamma)}{R}\right)|\sin(\gamma)|\right\rangle, \qquad (3)$$

where the angular brackets ($\langle\ldots\rangle$) represent averaging over the orientational distribution function ("ODF") of the particles. (Detailed analysis of the pairwise excluded volume [19,24] reveals that this approximation is justified provided: (i) $\lambda(\gamma) \ll L$, and: (ii) the dominant contributions to the excluded volume arise from configurations such that: $|\sin(\gamma)| \gg \lambda(\gamma)/L$ and $|\sin(\gamma)| \gg R/L$. We have assumed from the outset that the aspect ratio for the rods is much larger than unity, that is, that $L/R \gg 1$). Equation (3), along with the identification of the site occupation probability with particle volume fraction $\phi$, leads to the following expression for the vertex degree (or co-ordination number) for sites in the analogous lattice problem:

$$z = \left(\frac{4}{\pi}\right)\left(\frac{L}{R}\right)\left\langle\left(\frac{\lambda(\gamma)}{R}\right)|\sin(\gamma)|\right\rangle. \tag{4}$$

The foregoing lattice analogy has been augmented (Ref. [18]) in order to model effects due to particle clustering and inter-particle correlation, albeit in a strictly operational and mean-field manner. Variations in the degree of particle clustering are modeled by including a fraction of fully connected subgraphs in the lattice [25], and correlations between the states of occupancy of neighboring sites are treated within the Quasi-Chemical Approximation. Two parameters, denoted $p$ ($0 \le p \le 1$) and $K$ ($K > 0$), are introduced in order to quantify (operationally) deviations from a perfectly tree-like Bethe lattice with randomly occupied sites. The quantity $K$ is defined through the relationship: $K = p_{10}^2 / p_{00} p_{11}$, where $p_{ij}$ is the probability that a randomly chosen pair of adjacent sites have the occupancy states "$i$" and "$j$", where values of zero and unity denote vacant and occupied sites, respectively. A completely random, uncorrelated, state of site occupancies thus corresponds to $K = 1$. Larger (smaller) values of $K$ translate into situations where the number of inter-particle contacts is smaller (larger) than anticipated for a perfectly random arrangement, thereby providing an operational vehicle for capturing repulsive (attractive) inter-particle interactions. The quantity denoted $p$ equals the fraction of vertices that are in completely tree-like neighborhoods, and is related to the clustering coefficient, $C$, by way of: $C = (1-p)(1-2/z)$ for a monodisperse system. (The clustering coefficient is defined [26] as the probability that the members of a pair of randomly selected particles, each of which is known to be connected to another, third, particle, are also in contact with each other). In the parlance of this model, the cases: $p = 0$ and $p = 1$ correspond to maximal and minimal degrees of local clustering among the particles, respectively. In the latter ($p = 1$) case the lattice includes no closed loops and has a completely tree-like architecture. For fixed values of the rod aspect ratio ($L/R$), and given choices for the connectedness range $\lambda(\gamma)$ and the ODF for the rods, increases in $p$ and $K$ have the effect of lowering and elevating, respectively, the volume fraction $\phi_c$ at the percolation threshold.

In order to establish that given choices for the aspect ratio, volume fraction, ODF, $p$, $K$, and $\lambda(\gamma)$ correspond to the threshold for connectedness percolation, we first evaluate the auxiliary variable $x$ as the solution to [18,20]:

$$\phi(1-K)x^2 + Kx - K(1-\phi) = 0 \tag{5}$$

that satisfies: $0 \leq x \leq 1$. We next determine the value of the vertex degree, $z$, that ensures that prescriptions for $\phi$, $p$, $K$, $\lambda(\gamma)$, ODF, and aspect ratio correspond to the percolation threshold for these conditions by solving the following quadratic equation:

$$p(1-x)z^2 + \left[(1-p)(1-x)^2 - p(2-x)\right]z - (1-p)\left[(1-x)^2 + 1\right] = 0. \tag{6}$$

It can be readily verified that equations (5) and (6) always have unique solutions that belong to the ranges $0 \leq x \leq 1$ and $z > 1$, respectively. With the value of $z$ obtained from equations (5) and (6), application of equations (2) and (4) then allows determination of the conductivity $\sigma/\sigma_0$ for the given choice of parameters and ODF within the framework of the CPA. This procedure can be thought of as one in which the value of $\sigma/\sigma_0$ is chosen such that the resulting $\lambda(\gamma)$ (equation (2)) ensures that the dispersion of particles is precisely at the percolation threshold. Our strategy for an approximate evaluation of the orientational average in equation (4) for partially oriented rods is discussed next in Section 2.3.

**2.3. Averaging over the orientational distribution function:**

The ODF for the rods (denoted $f(\theta)$) is assumed to have azimuthal symmetry, and is normalized such that $\int_0^{2\pi} d\phi \int_0^{\pi} d\theta \sin(\theta) f(\theta) = 1$. For an explicit representation, we adopt the following model for $f(\theta)$ that was examined in Ref. [19]:

$$f(\theta) = a + b|\cos(\theta)|^m, \qquad (7)$$

where each of the parameters $b$ and $m$ are required to be greater than or equal to zero. Two of the three parameters $\{a,b,m\}$ introduced in equation (7) can be determined from (i) the normalization condition, which yields: $a + b/(m+1) = 1/4\pi$, and (ii) assignment of an average value to the orientational order parameter, $\langle S \rangle$, by way of:

$$\langle S \rangle = \langle P_2(\cos(\theta)) \rangle = \frac{4\pi bm}{(m+1)(m+3)}, \qquad (8)$$

where $P_2(x) = (3x^2 - 1)/2$ denotes the Legendre polynomial of second order. Requiring that the function $f(\theta)$ be positive-definite for any chosen value of $\langle S \rangle$ can be shown to restrict the range of allowed values of $m$ to satisfy:

$$\frac{3\langle S \rangle}{(1-\langle S \rangle)} \leq m. \qquad (9)$$

The latitude afforded by equation (9) in terms of the choice of *m* permits incorporation of variability in the standard deviation of the order parameter at a fixed value of $\langle S \rangle$:

$$\langle S^2 \rangle = \langle \{P_2(\cos(\theta))\}^2 \rangle = \frac{1}{5}\left[1 + \frac{2(2m+1)\langle S \rangle}{(m+5)}\right]. \tag{10}$$

The *ansatz* for $f(\theta)$ introduced in equation (7) thus provides sufficient flexibility to examine dependences upon both the first and second moments of *S*, and also suggests a pathway for the approximate analytical evaluation of the orientational averages that emerge from equations (2), (4), (5), and (6). It is straightforward (albeit tedious) to verify that for a fixed value of $\langle S \rangle$, the standard deviation in *S* increases monotonically with increase in the parameter *m*.

A method that has been frequently applied [27-29] towards the evaluation of orientational averages such as $\langle |\sin(\gamma)| \rangle$ and $\langle |\cos(\gamma)| \rangle$ in uniaxially aligned systems involves the following steps: (i) expressing the function to be averaged as a Fourier series in the even-ordered Legendre polynomials $\{P_{2n}\}$ in the variable $(\cos(\gamma))$, followed by (ii) use of the addition theorem for spherical harmonics, and finally (iii) integrating over the azimuthal angles by taking advantage of the symmetry of the problem. In order to apply this strategy to the present situation, we adopt the following approximate representation of the final term in the right-hand-side of equation (2) that is chosen to facilitate subsequent expansion in a Fourier-Legendre series:

$$\ln(\alpha^2 + \sin^2(\gamma)) \approx h_1 + h_2 \{|\sin(\gamma)|\}^{h_3}, \tag{11}$$

where $\alpha^2 = 4\pi R\xi/L^2$, and:

$$h_1 = \ln(\alpha^2), \; h_2 = \ln\left(1+\frac{1}{\alpha^2}\right), \text{ and : } h_3 = \frac{2\ln\left[\frac{\ln(1+1/\alpha^2)}{\ln(1+1/2\alpha^2)}\right]}{\ln(2)}. \tag{12}$$

Taken together with equation (12), the right-hand-side of equation (11) is constructed to accurately reproduce the values of the left-hand-side of equation (11) for $\gamma=0$, $\gamma=\pi/4$, and $\gamma=\pi/2$, and to also be (correctly) a monotonically increasing function of $\gamma$ over the interval: $0 \leq \gamma \leq \pi/2$. Equations (11) and (12) constitute a technical approximation that is intended to enable further analytical progress, and that can be expected to include the most important qualitative effects arising out of anisotropy in the tunneling probability. A comparison of the left and right hand sides of equation (11) is provided in figure 1 for values of the aspect ratio and tunneling decay length $\xi$ that are adopted in the model calculations presented in the following Section.

Equations (2), (4) and (11) lead to the following expression for $z$, which, together with equations (5) and (6), provides a model for the calculation of $\sigma/\sigma_0$ after the orientational averages are evaluated:

$$z = -\left(\frac{L}{\pi R}\right)\left(\frac{\xi}{R}\right)\left[\left(\ln\left(\frac{\sigma}{\sigma_0}\right)+h_1\right)\langle|\sin(\gamma)|\rangle + h_2\langle|\sin(\gamma)|^{(1+h_3)}\rangle\right], \tag{13a}$$

or, equivalently:

$$\ln\left(\frac{\sigma}{\sigma_0}\right) = -h_1 - \frac{h_2\langle|\sin(\gamma)|^{1+h_3}\rangle}{\langle|\sin(\gamma)|\rangle} - \frac{\pi z(R/L)(R/\xi)}{\langle|\sin(\gamma)|\rangle}. \tag{13b}$$

The Fourier-Legendre expansions of relevance to equation (13) are [30]:

$$|\sin(\gamma)|^q = \sum_{n=0}^{\infty} d_{2n;q} P_{2n}(\cos(\gamma)), \tag{14}$$

with:

$$d_{2n;q} = \frac{(4n+1)\pi \left[\Gamma\left(1+\frac{q}{2}\right)\right]^2}{2\Gamma\left(\frac{3}{2}+\frac{q}{2}+n\right)\Gamma\left(1+\frac{q}{2}-n\right)\Gamma(n+1)\Gamma\left(\frac{1}{2}-n\right)}, \tag{15}$$

where $\Gamma(x)$ denotes the Gamma function. Application of the addition theorem for spherical harmonics to each term on the right-hand-side of equation (14) followed by integration over the azimuthal angles leads to:

$$\langle |\sin(\gamma)|^q \rangle = 4\pi^2 \sum_{n=0}^{\infty} d_{2n;q} [I_{2n}]^2, \tag{16}$$

where:

$$I_{2n} = \int_0^{\pi} d\theta \sin(\theta) f(\theta) P_{2n}(\cos(\theta)). \tag{17}$$

Equations (7) and (17) yield [30]:

$$I_0 = \frac{1}{2\pi}, I_2 = \frac{\langle S \rangle}{2\pi}, \text{ and: } I_4 = \frac{(m-2)\langle S \rangle}{2\pi(m+5)}, \tag{18}$$

together with the following recurrence relation that generates the higher-order terms in the sequence of $\{I_{2n}\}$:

$$\frac{I_{2n+2}}{I_{2n}} = \frac{(m/2-n)}{(n+m/2+3/2)}, \text{ for } n \geq 2. \tag{19}$$

These results simplify considerably in the limit that $m \to \infty$. In this limiting case, equation (16) reduces to:

$$\langle|\sin(\gamma)|^q\rangle = \frac{\sqrt{\pi}\left[1-\langle S\rangle^2\right]\Gamma(1+q/2)}{2\Gamma(3/2+q/2)}, \text{ when } m\to\infty. \tag{20}$$

Equations (5), (6), (13), (15), (16), (18), and (19) provide a route towards calculating the conductivity as a function of the volume fraction, degree of clustering and correlation among particles, and extent of orientational ordering.

### 2.4. EMA conductivity:

To facilitate comparison of the conductivity obtained from the method described in the previous sections with that evaluated within the EMA of Ref. [17], we report below the self-consistent EMA equation the solution to which provides $\sigma/\sigma_0$:

$$\frac{\xi L}{\pi R^2}\varphi\left\langle\begin{array}{c}\sin(\gamma)\left\{1+\dfrac{\alpha^2(\sigma/\sigma_0)}{\left(1+(\sigma/\sigma_0)\sin^2(\gamma)\right)}\right\}\ln\left[1+\dfrac{1}{(\sigma/\sigma_0)(\alpha^2+\sin^2(\gamma))}\right]\\ \\ -\dfrac{\alpha^2}{\left(1+(\sigma/\sigma_0)\sin^2(\gamma)\right)\sin(\gamma)}\ln\left(\dfrac{\alpha^2+\sin^2(\gamma)}{\alpha^2}\right)\end{array}\right\rangle = 1, \tag{21}$$

where the angular brackets represent the average over the ODF given in equation (7). The pre-factor in front of the angular brackets has been normalized so as to reproduce the CPA conductivity for homogeneous distributions of rods in the dilute limit, as discussed in Ref. [17]. We note that, in contrast to the tunneling conductance of equation (1), the above expression also accounts for configurations of rods for which the centers of the particles are not located on the line of shortest distance between the rod axes. However, as we shall find in the next section where the solution of equation (21) is compared with the CPA conductivity obtained from equation (1), accounting for these configurations has only a weak effect upon the calculated value of the conductivity. Our comparison of results from (i) the lattice analogy for percolation combined with the CPA, and (ii) from the EMA, will be restricted to the case of fully unclustered local configurations and random, uncorrelated particle locations, that is to the choice: $p = K = 1$ within the former approach, as the EMA formulated in Ref. [17] does not explicitly include effects arising out of the degree of local clustering or correlation.

## 3. Results:

This Section presents results for the dependence of the conductivity upon particle volume fraction and the degree of orientational ordering, evaluated for the model discussed in Section 2. Results for the conductivity are in each case expressed in terms of the dimensionless quantity $\sigma/\sigma_0$ that (within the CPA) denotes the critical value of the tunneling conductance corresponding to a choice of inter-surface separation which enforces the percolation threshold condition for given values of the remaining parameters. All of the calculations discussed in this work are performed for rods with an aspect ratio equal to 200 (that is, $L/R = 200$) and the tunneling decay length $\xi$ is assigned the value $\xi = 0.2R$. For these parameters, the solid and broken curves in figure 1 display the left- and right-hand-sides of equation (11), respectively, as functions of the relative angle between a pair of rods. The approximation of equation (11), while qualitatively not unreasonable, results in an *overestimate* for small values of $\gamma$. Scrutiny of equation (2) shows that this implies an *underestimate* of the mutual conductance between a pair of rods in the same regime (near-parallel orientations), and thus, our subsequent findings for the conductivity can be anticipated to be lower bounds (within the present model), especially for large values of the order parameter $\langle S \rangle$.

The dependence of $\sigma/\sigma_0$ upon the mean orientational order parameter $\langle S \rangle$ is examined in figures 2, 3, 4, and 5. In each instance, enhancement in the degree of alignment (quantified by $\langle S \rangle$) leads to reduced conductivity when all other variables are held fixed. The increase in the typical inter-particle distance (resulting in lower conductivity) that accompanies stronger alignment dominates over the conductivity enhancement that arises due to the simultaneous reduction in the angle $\gamma$ between the rod axes. Conductivity is found to increase with the particle volume fraction $\phi$ (figure 2). Results from the CPA-based and EMA-based approaches are compared in figure 3, where reasonably good agreement is observed between these different routes of calculation. The conductivity decreases when inter-particle correlations are such as to reduce the number of inter-particle contacts (figure 4), an effect that our lattice percolation model describes through an increase in the value of $K$. The results displayed in figure 5 show that clustering among the particles (smaller value of $p$) also lowers the conductivity. Interestingly, we observe that in all cases, $\sigma/\sigma_0$ is highly sensitive to the standard deviation of the order parameter (controlled by our choice for $m$, equation (10)). Increasing values of $m$ reflect an ODF that is more strongly peaked towards parallel orientations, implying larger separations between particle surfaces and a concomitant drop in the conductivity.

The dependence of $\sigma/\sigma_0$ upon the volume fraction is explored in figures 6, 7, and 8. The conductivity is found to be an increasing function of $\phi$ when all of the other parameters are held constant. Greater degrees of uniaxial alignment (figure 6) or increased particle clustering (figure 8) lead to reductions in $\sigma/\sigma_0$, and notable sensitivity to the variance in $S$ is observable. Comparison with the volume fraction dependence predicted by the EMA-based approach (equation (21), figure 7) reveals good agreement (especially for small volume fractions) and similar variability in the results as functions of $\sigma_S$). (The symbol $\sigma_S$ denotes the standard deviation in the order parameter, not to be confused with conductivity). The effective conductivity of the composite is strongly dependent upon details of the microstructure, and may vary significantly even under conditions of fixed volume fraction $\phi$ and mean order parameter $\langle S \rangle$.

In closing, we consider an approximation that decouples effects arising from the two distinct sources of anisotropy, namely, anisotropy in the tunneling probability and that modeled by an anisotropic ODF. Towards this end, we model the denominator of the right-hand-side of equation (1) by way of the isotropic, $\gamma$-independent, quantity introduced below:

$$\ln(\alpha^2 + \sin^2(\gamma)) \to h_1 + \frac{h_2 \langle |\sin(\gamma)|^{1+h_3} \rangle_{iso}}{\langle |\sin(\gamma)| \rangle_{iso}} = h_1 + \frac{2h_2 \Gamma\left(\frac{3}{2} + \frac{h_3}{2}\right)}{\sqrt{\pi}\, \Gamma\left(2 + \frac{h_3}{2}\right)}, \tag{22}$$

where $h_1$, $h_2$, and $h_3$ are defined as in equation (12) and the averages are now taken over an *isotropic* orientational distribution. Use of the representation in equation (22), as opposed to that in equation (11), leads to the following analogue to equation (13b):

$$\ln\left(\frac{\sigma}{\sigma_0}\right) = -h_1 - \frac{\pi z (R/L)(R/\xi)}{\langle|\sin(\gamma)|\rangle} - \frac{2h_2 \Gamma\left(\frac{3}{2} + \frac{h_3}{2}\right)}{\sqrt{\pi}\,\Gamma\left(2 + \frac{h_3}{2}\right)}. \quad (23)$$

The approximation introduced in equations (22) and (23), when used in conjunction with the rest of our model, yields identical results to those obtained from equations (11) and (13b) for the case of an isotropic distribution and allows us to separate effects of tunneling anisotropy from those that originate from an anisotropic ODF. Interestingly enough, for the choice of ODF in equation (7), equations (13b) and (23) also become identical in the limit that $m \to \infty$ for any value of $\langle S \rangle$, that is:

$$\underset{m \to \infty}{\text{Lt}} \frac{\langle|\sin(\gamma)|^{1+h_3}\rangle}{\langle|\sin(\gamma)|\rangle} = \frac{\langle|\sin(\gamma)|^{1+h_3}\rangle_{iso}}{\langle|\sin(\gamma)|\rangle_{iso}} = \frac{2\Gamma\left(\frac{3}{2} + \frac{h_3}{2}\right)}{\sqrt{\pi}\,\Gamma\left(2 + \frac{h_3}{2}\right)}. \quad (24)$$

Equations (13b) and (23) therefore also lead to the same results for arbitrary values of $\langle S \rangle$ in the limit that $m \to \infty$. Figure 9 compares results from calculations based upon equations (11) and (13) (solid lines: both tunneling and orientational anisotropy are accounted for) with those that employ equations (22) and (23) (dashed lines: only orientational anisotropy is included). Accounting for anisotropy in the tunneling probability leads to enhancement in the predicted conductivity, which is reasonable in light of the fact that equation (1) implies higher conductances between pairs of parallel versus perpendicular rods. However, this effect is insufficient (within the present approach and for our choice of ODF) to reverse the monotonic decline in $\sigma/\sigma_0$ with increasing $\langle S \rangle$ in case the other variables that (operationally) describe the microstructure (*p* and *K*) are assigned constant, $\langle S \rangle$-independent values.

## 4. Summary and Conclusions:

An analogy between lattice and continuum percolation that was previously applied towards estimating $\phi_c$ for partially oriented, polydisperse systems of rods has been used in conjunction with the CPA and a recent model for the tunneling anisotropy to describe the conductance in dispersions of aligned particles. Approximations are introduced that enable a treatment that is largely analytical and that includes effects due to particle clustering. Our choice of orientational distribution permits modeling both the mean value and standard deviation of the order parameter *S*.

In each case we have examined, the conductivity decreases monotonically with increasing $\langle S \rangle$ for fixed values of the remaining variables. This finding is similar to that reported in a recent EMA-based investigation [17]. However, the present work suggests that the conductivity may display a strong dependence upon the variance in the order parameter, in addition to the degree of particle clustering. Observations of non-monotonic variation [13,14] in the conductivity with $\langle S \rangle$ for fixed particle volume fractions could conceivably arise from subtle alterations in inter-particle clustering and the standard deviation in *S* as the mean degree of alignment is increased.

**Figure Captions:**

**Figure 1.** The solid and broken lines show, respectively, the left and right hand sides of equation (11) as functions of the angle $\gamma$. In each case $L = 200\ R$ and $\xi = 0.2\ R$. The parameters $\{h_1, h_2, h_3\}$ for use in equation (11) are calculated according to equation (12).

**Figure 2.** The logarithm (to base ten) of the conductivity is shown as a function of the mean orientational order parameter for rods for which $L = 200\ R$ and $\xi = 0.2\ R$. For all of the curves in this figure, $p = K = 1$. The pair of upper curves, solid and broken, represent results when $\phi = 0.05$ and $m = m_{MIN} = 3\langle S \rangle / (1 - \langle S \rangle)$ and $m \to \infty$, respectively. The pair of lower curves, solid and broken, represent results when $\phi = 0.01$ and $m = m_{MIN} = 3\langle S \rangle / (1 - \langle S \rangle)$ and $m \to \infty$, respectively.

**Figure 3.** The logarithm (to base ten) of the conductivity is shown as a function of the mean orientational order parameter for rods for which $L = 200\ R$ and $\xi = 0.2\ R$. For all of the curves in this figure, $\phi = 0.01$. The pair of solid curves, upper and lower, represent results from the percolation-based approach within the lattice analogy when $p = K = 1$ and $m = m_{MIN} = 3\langle S \rangle / (1 - \langle S \rangle)$ (upper) and $m \to \infty$ (lower), respectively. The pair of broken (dashed) curves, upper and lower, represent results obtained from the EMA-based approach for the same ODF (equation (21)) and with $m = m_{MIN} = 3\langle S \rangle / (1 - \langle S \rangle)$ (upper) and $m \to \infty$ (lower), respectively.

**Figure 4.** The logarithm (to base ten) of the conductivity is shown as a function of the mean orientational order parameter for rods for which $L = 200\ R$ and $\xi = 0.2\ R$. For all of the curves in this figure, $p = 1$ and $\phi = 0.05$. The pair of upper curves, solid and broken,

represent results when $K = 1$ and $m = m_{MIN} = 3\langle S \rangle / (1 - \langle S \rangle)$ and $m \to \infty$, respectively. The pair of lower curves, solid and broken, represent results when $K = 2$ and $m = m_{MIN} = 3\langle S \rangle / (1 - \langle S \rangle)$ and $m \to \infty$, respectively.

**Figure 5.** The logarithm (to base ten) of the conductivity is shown as a function of the mean orientational order parameter for rods for which $L = 200\ R$ and $\xi = 0.2\ R$. For all of the curves in this figure, $K = 1$ and $\phi = 0.05$. The pair of upper curves, solid and broken, represent results when $p = 1$ and $m = m_{MIN} = 3\langle S \rangle / (1 - \langle S \rangle)$ and $m \to \infty$, respectively. The pair of lower curves, solid and broken, represent results when $p = 0.25$ and $m = m_{MIN} = 3\langle S \rangle / (1 - \langle S \rangle)$ and $m \to \infty$, respectively.

**Figure 6.** The logarithm (to base ten) of the conductivity is shown as a function of the volume fraction $\phi$ for fixed values of the mean orientational order parameter for rods for which $L = 200\ R$ and $\xi = 0.2\ R$. For all of the curves in this figure, $p = K = 1$. The dotted line (uppermost curve) displays results for an isotropic system, $\langle S \rangle = 0$. The pair of upper curves, solid and broken (dashed), represent results when $\langle S \rangle = 0.4$ and $m = m_{MIN} = 3\langle S \rangle / (1 - \langle S \rangle)$ and $m \to \infty$, respectively. The pair of lower curves, solid and broken (dashed), represent results when $\langle S \rangle = 0.8$ and $m = m_{MIN} = 3\langle S \rangle / (1 - \langle S \rangle)$ and $m \to \infty$, respectively.

**Figure 7.** The logarithm (to base ten) of the conductivity is shown as a function of the volume fraction $\phi$ for fixed values of the mean orientational order parameter for rods for which $L = 200\ R$ and $\xi = 0.2\ R$. For all of the calculations performed using the percolation-based approach within the lattice analogy, $p = K = 1$. The pair of dotted lines,

upper and lower, displays results for an isotropic system for which $\langle S \rangle = 0$ calculated from the EMA-based (equation (21)) (upper) and the percolation-based (lower) approaches, respectively. The pair of solid curves, upper and lower, represent results from the percolation-based approach when $\langle S \rangle = 0.8$ and $m = m_{MIN} = 3\langle S \rangle / (1-\langle S \rangle)$ (upper) and $m \to \infty$ (lower), respectively. The pair of broken (dashed) curves, upper and lower, represent results from the EMA-based (equation (21)) approach when $\langle S \rangle = 0.8$ for the same ODF and with $m = m_{MIN} = 3\langle S \rangle / (1-\langle S \rangle)$ (upper) and $m \to \infty$ (lower), respectively.

**Figure 8.** The logarithm (to base ten) of the conductivity is shown as a function of the volume fraction $\phi$ for fixed values of the mean orientational order parameter for rods for which $L = 200\ R$ and $\xi = 0.2\ R$. For all of the curves in this figure, $K = 1$. The pair of dotted lines (uppermost curves) display results for isotropic systems, $\langle S \rangle = 0$, for which $p = 1$ and $p = 0.1$ for the upper and lower dotted curves, respectively. The pair of solid curves represent results when $\langle S \rangle = 0.8$ and $m = m_{MIN} = 3\langle S \rangle / (1-\langle S \rangle)$, for $p = 1$ and $p = 0.1$ for the upper and lower solid curves, respectively. The pair of broken (dashed) curves represent results when $\langle S \rangle = 0.8$ and $m \to \infty$, for $p = 1$ and $p = 0.1$ for the upper and lower dashed curves, respectively.

**Figure 9.** The logarithm (to base ten) of the conductivity is shown as a function of the mean orientational order parameter for rods for which $L = 200\ R$ and $\xi = 0.2\ R$. For all of the curves in this figure, $p = K = 1$ and $\phi = 0.025$. The solid and broken lines represent results calculated from equations (11) and (13) (both tunneling and orientational

anisotropy included), and from equations (22) and (23) (orientational anisotropy alone), respectively. For each set of curves, from uppermost to lowermost, $m = m_{MIN} = 3\langle S \rangle / (1 - \langle S \rangle)$, $m = 10 m_{MIN}$, and $m \to \infty$, respectively. The lowermost solid and broken curves coincide with each other (equation (24)).

**Figure 1**

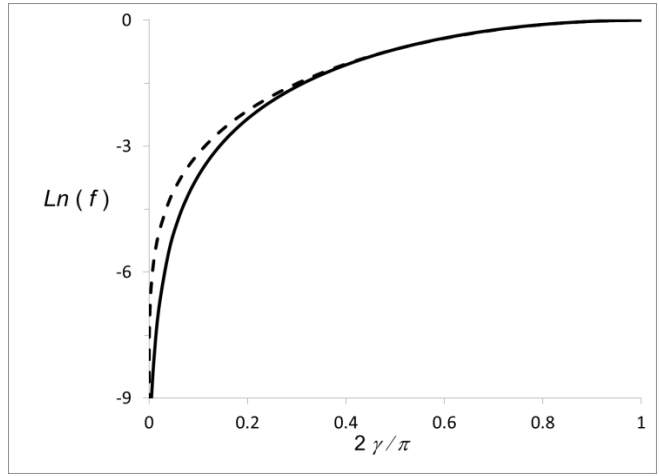

**Figure 1.** The solid and broken lines show, respectively, the left and right hand sides of equation (11) as functions of the angle $\gamma$. In each case $L = 200\,R$ and $\xi = 0.2\,R$. The parameters $\{h_1, h_2, h_3\}$ for use in equation (11) are calculated according to equation (12).

**Figure 2**

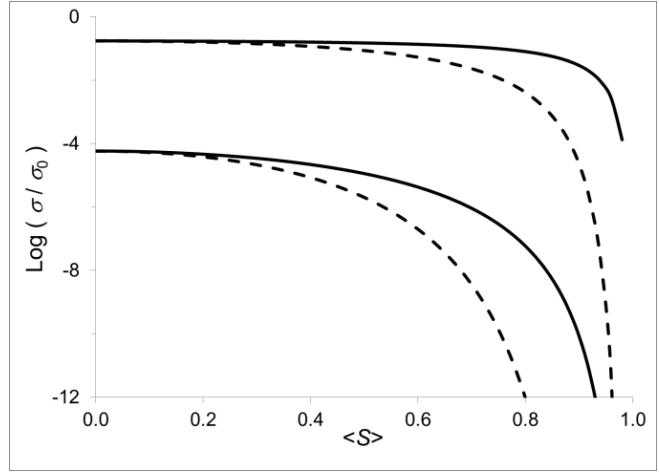

**Figure 2.** The logarithm (to base ten) of the conductivity is shown as a function of the mean orientational order parameter for rods for which $L = 200\ R$ and $\xi = 0.2\ R$. For all of the curves in this figure, $p = K = 1$. The pair of upper curves, solid and broken, represent results when $\phi = 0.05$ and $m = m_{MIN} = 3\langle S\rangle/(1-\langle S\rangle)$ and $m \to \infty$, respectively. The pair of lower curves, solid and broken, represent results when $\phi = 0.01$ and $m = m_{MIN} = 3\langle S\rangle/(1-\langle S\rangle)$ and $m \to \infty$, respectively.

**Figure 3**

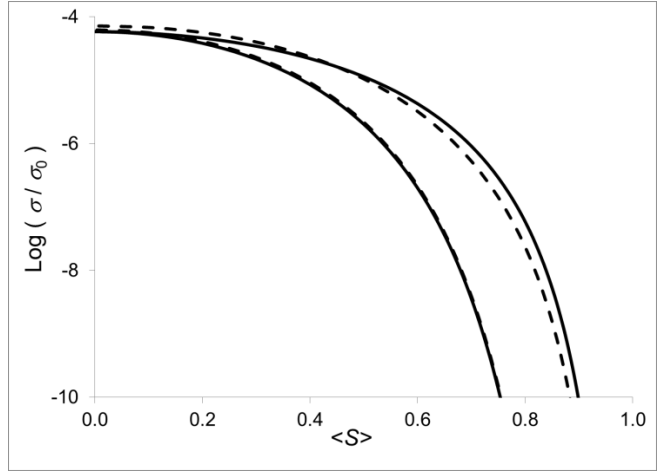

**Figure 3.** The logarithm (to base ten) of the conductivity is shown as a function of the mean orientational order parameter for rods for which $L = 200\,R$ and $\xi = 0.2\,R$. For all of the curves in this figure, $\phi = 0.01$. The pair of solid curves, upper and lower, represent results from the percolation-based approach within the lattice analogy when $p = K = 1$ and $m = m_{MIN} = 3\langle S \rangle / (1 - \langle S \rangle)$ (upper) and $m \to \infty$ (lower), respectively. The pair of broken (dashed) curves, upper and lower, represent results obtained from the EMA-based approach for the same ODF (equation (21)) and with $m = m_{MIN} = 3\langle S \rangle / (1 - \langle S \rangle)$ (upper) and $m \to \infty$ (lower), respectively.

**Figure 4**

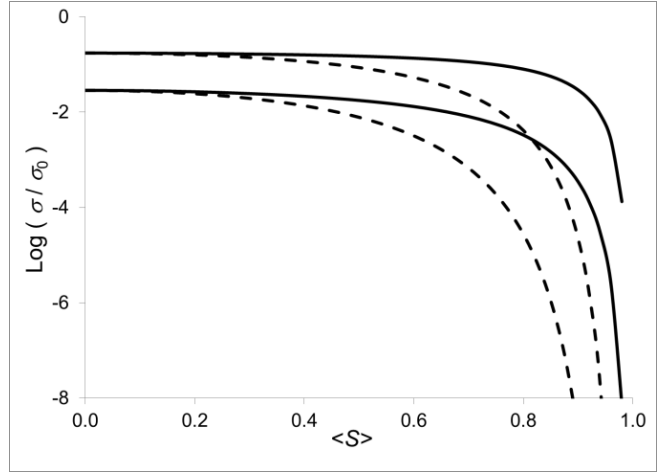

**Figure 4.** The logarithm (to base ten) of the conductivity is shown as a function of the mean orientational order parameter for rods for which $L = 200\ R$ and $\xi = 0.2\ R$. For all of the curves in this figure, $p = 1$ and $\phi = 0.05$. The pair of upper curves, solid and broken, represent results when $K = 1$ and $m = m_{MIN} = 3\langle S \rangle / (1 - \langle S \rangle)$ and $m \to \infty$, respectively. The pair of lower curves, solid and broken, represent results when $K = 2$ and $m = m_{MIN} = 3\langle S \rangle / (1 - \langle S \rangle)$ and $m \to \infty$, respectively.

**Figure 5**

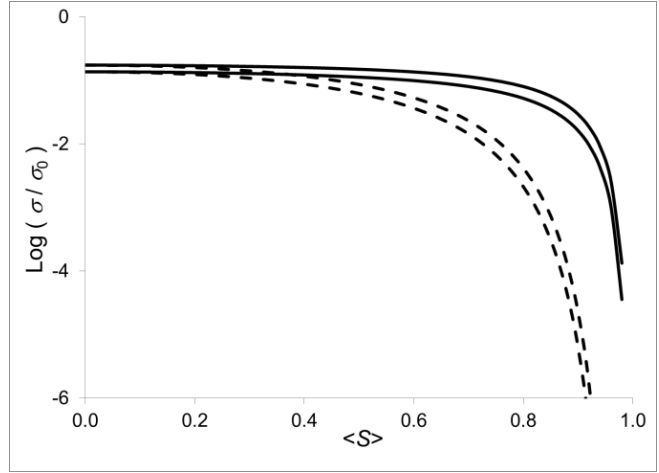

**Figure 5.** The logarithm (to base ten) of the conductivity is shown as a function of the mean orientational order parameter for rods for which $L = 200\ R$ and $\xi = 0.2\ R$. For all of the curves in this figure, $K = 1$ and $\phi = 0.05$. The pair of upper curves, solid and broken, represent results when $p = 1$ and $m = m_{MIN} = 3\langle S \rangle/(1-\langle S \rangle)$ and $m \to \infty$, respectively. The pair of lower curves, solid and broken, represent results when $p = 0.25$ and $m = m_{MIN} = 3\langle S \rangle/(1-\langle S \rangle)$ and $m \to \infty$, respectively.

**Figure 6**

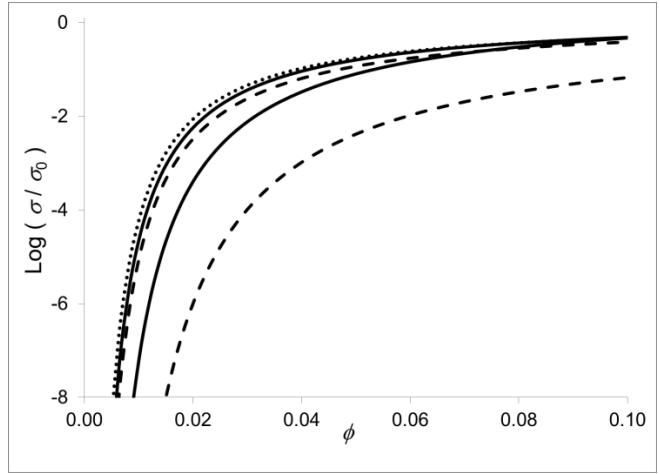

**Figure 6.** The logarithm (to base ten) of the conductivity is shown as a function of the volume fraction $\phi$ for fixed values of the mean orientational order parameter for rods for which $L = 200\,R$ and $\xi = 0.2\,R$. For all of the curves in this figure, $p = K = 1$. The dotted line (uppermost curve) displays results for an isotropic system, $\langle S \rangle = 0$. The pair of upper curves, solid and broken (dashed), represent results when $\langle S \rangle = 0.4$ and $m = m_{MIN} = 3\langle S \rangle/(1-\langle S \rangle)$ and $m \to \infty$, respectively. The pair of lower curves, solid and broken (dashed), represent results when $\langle S \rangle = 0.8$ and $m = m_{MIN} = 3\langle S \rangle/(1-\langle S \rangle)$ and $m \to \infty$, respectively.

**Figure 7**

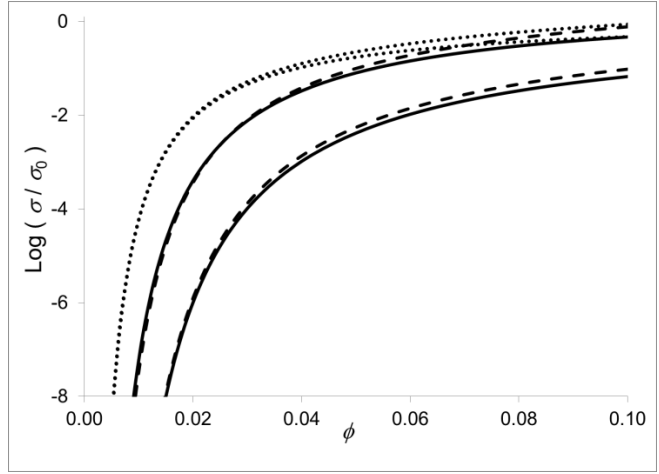

**Figure 7.** The logarithm (to base ten) of the conductivity is shown as a function of the volume fraction $\phi$ for fixed values of the mean orientational order parameter for rods for which $L = 200\, R$ and $\xi = 0.2\, R$. For all of the calculations performed using the percolation-based approach within the lattice analogy, $p = K = 1$. The pair of dotted lines, upper and lower, displays results for an isotropic system for which $\langle S \rangle = 0$ calculated from the EMA-based (equation (21)) (upper) and the percolation-based (lower) approaches, respectively. The pair of solid curves, upper and lower, represent results from the percolation-based approach when $\langle S \rangle = 0.8$ and $m = m_{MIN} = 3\langle S \rangle/(1-\langle S \rangle)$ (upper) and $m \to \infty$ (lower), respectively. The pair of broken (dashed) curves, upper and lower, represent results from the EMA-based (equation (21)) approach when $\langle S \rangle = 0.8$ for the same ODF and with $m = m_{MIN} = 3\langle S \rangle/(1-\langle S \rangle)$ (upper) and $m \to \infty$ (lower), respectively.

**Figure 8**

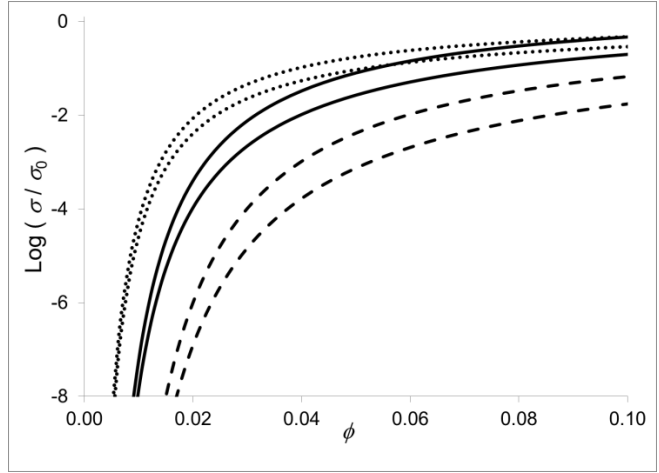

**Figure 8.** The logarithm (to base ten) of the conductivity is shown as a function of the volume fraction $\phi$ for fixed values of the mean orientational order parameter for rods for which $L = 200\ R$ and $\xi = 0.2\ R$. For all of the curves in this figure, $K = 1$. The pair of dotted lines (uppermost curves) display results for isotropic systems, $\langle S \rangle = 0$, for which $p = 1$ and $p = 0.1$ for the upper and lower dotted curves, respectively. The pair of solid curves represent results when $\langle S \rangle = 0.8$ and $m = m_{MIN} = 3\langle S \rangle / (1 - \langle S \rangle)$, for $p = 1$ and $p = 0.1$ for the upper and lower solid curves, respectively. The pair of broken (dashed) curves represent results when $\langle S \rangle = 0.8$ and $m \to \infty$, for $p = 1$ and $p = 0.1$ for the upper and lower dashed curves, respectively.

**Figure 9**

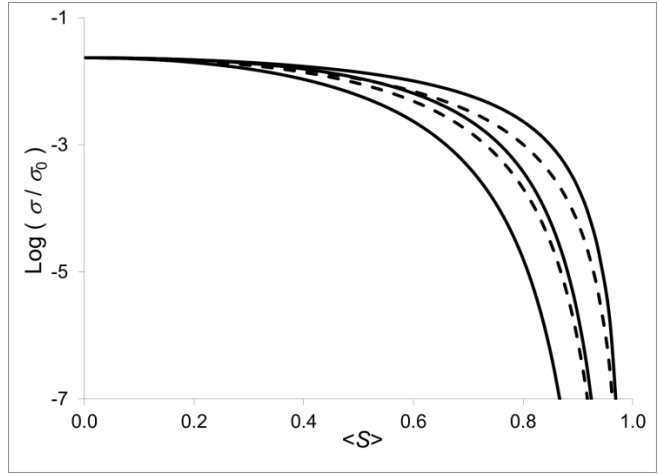

**Figure 9.** The logarithm (to base ten) of the conductivity is shown as a function of the mean orientational order parameter for rods for which $L = 200\ R$ and $\xi = 0.2\ R$. For all of the curves in this figure, $p = K = 1$ and $\phi = 0.025$. The solid and broken lines represent results calculated from equations (11) and (13) (both tunneling and orientational anisotropy included), and from equations (22) and (23) (orientational anisotropy alone), respectively. For each set of curves, from uppermost to lowermost, $m = m_{MIN} = 3\langle S \rangle / (1 - \langle S \rangle)$, $m = 10 m_{MIN}$, and $m \to \infty$, respectively. The lowermost solid and broken curves coincide with each other (equation (24)).